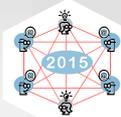

# uIP Support for the Network Simulation Cradle

Michael Kirsche and Roman Kremmer
Computer Networks and Communication Systems Group
Brandenburg University of Technology Cottbus-Senftenberg, Germany
eMail: {michael.kirsche, roman.kremmer}@b-tu.de

*Abstract*—We introduce the ongoing integration[1] of Contiki's uIP stack into the OMNeT++ port of the Network Simulation Cradle (NSC). The NSC utilizes code from real world stack implementations and allows for an accurate simulation and comparison of different TCP/IP stacks and a validation of thereby connected simulation models. uIP(v6) provides resource-constrained devices with an RFC-compliant TCP/IP stack and promotes the use of IPv6 in the vastly growing field of Internet of Things scenarios. This work-in-progress report discusses our motivation to integrate uIP into the NSC, our chosen approach and possible use cases for the simulation of uIP in OMNeT++.

## I. INTRODUCTION

TCP/IP stacks are the global enablers of world-wide interconnection and communication of devices and computers. A wide range of stack implementations is available today: from full-scale stacks in Windows and Linux to small-scale stacks for embedded systems like *lightweight IP* (lwIP) [1] and even smaller stacks like *micro IP* (uIP) [1] for resource-constrained devices with 8-bit microcontrollers. Using IP(v6) [2] as the bridging protocol facilitates a seamless interconnection of both resource-constrained and non-constrained devices over the Internet (condensed under the label *Internet of Things*).

Adaptation protocols like 6LoWPAN [3], routing protocols like RPL, and application protocols like CoAP and XMPP [4] use the functionalities of IP. A thorough testing of IP stacks is thus necessary; not only self-contained but also in comparison and interconnection with other deployed stacks and protocols. Testing uIP is usually done both via real-life implementations in testbeds and via simulations with the Contiki simulator Cooja [5]. Contiki is an operating system for embedded (wireless) devices, which contains an implementation of uIP with IPv4 and IPv6 capabilities [6]. Cooja is Contiki's own simulator, enabling a time-accurate simulation of Contiki source code through the use of Java Native Interfaces (JNI) and microcontroller simulators like MSPsim [7] in combination with an event-based simulation core and abstract channel models.

While practical testing is essential before any type of rollout, it is typically only feasible for small scale applications. Testing large scale Internet of Things scenarios, especially when cooperating with Internet-based systems, is important yet hard to manage with large numbers of sensors and actuators, different backbone networks and various Internet-based systems working together. For such cases, simulation is still a standard approach to acquire impressions on run-time characteristics and possible bottlenecks. Testing one IP stack against another is also mandatory to ensure interoperability and to assess and compare a stack's performance (refer to [8]).

[1]Source code available online: https://github.com/michaelkirsche/uip4nsc

The Cooja simulator provides accurate results for code execution times and energy consumption of Contiki-based sensor networks. But it lacks the universality of generic frameworks like OMNeT++ and the ability to simulate diverse communication protocols and technologies. Cooja does allow for a socket-based connection to other computers and real networks with the help of virtual network interfaces and the Serial Line Internet Protocol (SLIP), which can be used to connect nodes in Cooja to other networks and hence emulate border router scenarios where Contiki nodes communicate with non-Contiki nodes over a gateway connection. This enables the emulation of simple Internet of Things scenarios, which strive for an interconnection with Internet-based systems and communication over the Internet. The set-up and management of other communication stacks, devices and protocols outside of the Contiki world lies beyond the scope of Cooja. We think that a universal simulation framework like OMNeT++ provides a more generic environment to simulate scenarios that go above and beyond plain Contiki-based sensor networks. Examining IP stacks from the IoT world is also an important task, we therefore aim to extend INET with uIP support to extend OMNeT++'s applicability for IoT simulations. Extending and using OMNeT++ / INET provides the flexibility and model support to simulate IoT applications that interact over gateways, bridges and backbone networks, while supporting various communication technologies and keeping everything in a single controllable environment.

To provide an adequate simulation of IoT applications, we require accurate simulation models for used protocols. Modeling a complete TCP/IP stack from scratch and validating it afterwards is tedious and failure-prone work. [9] showed that a direct integration of a real-world TCP/IP stack into OMNeT++ is possible. One of the main problems for this type of integration is the constant maintenance necessary to adapt to latest code changes. A different integration approach was proposed by Samuel Jansen with his Network Simulation Cradle (NSC) [10]. The NSC wraps real world stacks to make them available to network simulators with lesser need for maintenance. Jansen's original work for ns-2 has proven to be extendable to OMNeT++ and additional TCP/IP stacks like lwIP. It facilitates a validation of simulated stacks [11], [12]. We therefore favor to integrate uIP into the NSC to simulate uIP-based IoT scenarios with OMNeT++.

The remainder of this report is structured as follows. Further motivational aspects for an integration of uIP into the NSC as well as background information are presented in section II. An overview of our integration approach and validation aspects are introduced in section III. A discussion of possible application scenarios in section IV and a summary of ongoing work in section V complete the report.

## II. uIP and the Network Simulation Cradle (NSC)

The investigation of network protocols like uIP is often conducted via simulations to study their behavior and reactions to different parameter settings in large scale environments. However, simulation results are only reliable when simulation models are verified to emulate the real protocol behavior. Validated implementations of protocols in simulators are hence an important precondition for meaningful simulations, while missing validations cast doubts over achieved results.

The Network Simulation Cradle was developed to meet the requirements of an accurate but still feasible simulation of TCP/IP communication stacks. A comparison [10] of common simulation models with testbed measurements showed distinct differences between real world stacks and simulation models and even between different real world stacks interacting with each other. To achieve a realistic behavior in simulations, the NSC utilizes real network stacks from common and available implementations instead of focusing on enhancing already abstracted simulation models. The NSC itself was validated[2] by testing conducted simulations against laboratory setups and a network testbed to ensure a realistic emulation of both protocol behavior (validated by checking the sequence of generated packets) and protocol performance (by comparing achieved goodputs in both simulation and testbed).

In an earlier work by Bless and Doll [9], the motivation was to build on an already validated and proven implementation of the TCP/IP stack as a simulation tool by porting the FreeBSD stack to OMNeT++ and avoid a tedious validation process afterwards. The porting process resulted in several manual code changes, which made maintenance and keeping the stack up-to-date very time-consuming. In addition, the possibility of human errors during the porting process could not be excluded completely. The NSC was developed with the precondition to avoid manual changes of the source code and to facilitate the integration and maintenance of different available TCP/IP stacks. In a nutshell, the approach is to compile the extracted network code of the TCP/IP stacks with support functions and a simulator interface into a shared library, which in turn is used by and linked against the simulation framework.

Running multiple instances of a network stack on a single machine is achieved with the help of a parser, which automatically modifies global variables and their references to enable a static virtualization of C source code [13]. This is necessary because network stacks in operating systems are not designed to allow for multiple instances running at the same time as stacks are simply not re-entrant [14, Sec. 3.4]. Each instance of a stack needs his own version of all global variables to run in parallel. To achieve this, one could compile the stack as a shared library and load multiple instances of it or create a new process for every simulated instance. Both approaches lack efficiency and scalability for simulations of large numbers of TCP/IP instances. Static virtualization by filtering the preprocessed C code is used instead to fulfill NSC's goal of adding as little overhead as possible and refraining from manual code changes. A "globaliser parser" is used to programmatically replace all instances of global variables and all references to them in the C code [15]. The parser's basic premise is demonstrated in the following listings 1 and 2, taken from the uIP code.

Listing 1. Original uIP source code
```
struct uip_conn *uip_conn;  /* current connection */

void uip_process(uint8_t flag)
{ ...
  uip_conn = NULL;
```

Listing 2. After executing the globaliser parser
```
struct uip_conn *global_uip_conn[NUM_STACKS];

void uip_process(uint8_t flag)
{ ...
  global_uip_conn[get_stack_id()] = NULL;
```

Today, the NSC supports the Linux, FreeBSD, OpenBSD and the lwIP stack[3]. While initially developed for the ns-2 simulator, a later port made the NSC and thus supported stacks available for OMNeT++.

Our motivation to provide uIP support for the NSC arises from the fact that uIP was designed for a different class of systems, namely resource-constrained systems. Basic premise for the design was a very low memory footprint [16, Table 13.1] to facilitate the use of uIP on 8- and 16-bit microcontrollers with small (Kilobytes) RAM and ROM sizes. The design premises lead to various well-known performance issues. TCP packets, for example, are sent and acknowledged one at a time. The "one-packet-at-a-time" logic is derived from the fact that uIP uses only a single (and stack traversing) packet buffer to support even the smallest platforms. These and other constrains are documented for example in [1].

While uIP ⇌ uIP communication is not affected by these constraints, uIP ⇌ "larger" TCP/IP stack communication can be affected if, for example, the communication partner uses delayed acknowledgments as defined by RFC 1122 [17]. These delayed ACKs effectively constrain the packet transmission. A performance hack for this TCP throughput issue is provided in Contiki [18]. We want to compare uIP's performance, especially when uIP is used together with full-scale TCP/IP stacks from non-constrained Internet-based systems. We also want to evaluate other limitations of uIP in combination with MAC protocols like IEEE 802.15.4, such as the effects of different lower layer fragmentation values on application layer protocol performance. These and other use cases for a NSC with uIP support will be further discussed in section IV.

---

[2]NSC validation: http://www.wand.net.nz/~stj2/nsc/validation.html

[3]NSC stack status: http://www.wand.net.nz/~stj2/nsc/status.html

### III. INTEGRATION AND VALIDATION

The integration of Contiki's uIP stack follows the example of other supported stacks like the Linux or the lwIP stack. Files containing the network code are integrated into the build process without any modifications. References to unused system functions are implemented as stub functions. The simulator interface utilizes the uIP API while access functions for Contiki's MAC layer are reimplemented to redirect back to the simulator (OMNeT++) interface. Figure 1 depicts the basic principle of the integration with the according function and file names.

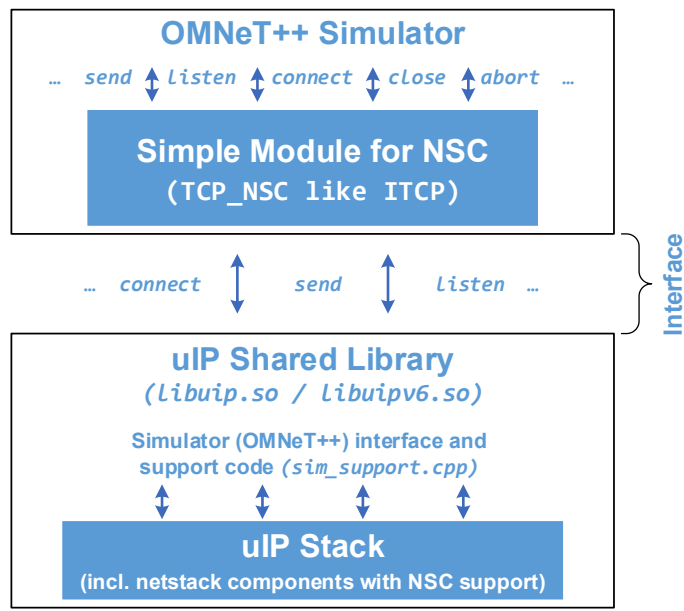

Fig. 1. Integration of uIP in the OMNeT++ NSC module and Interaction between OMNeT++ and uIP via the NSC

The *globaliser parser* changes global variables and their references automatically to arrays. It is integrated as part of the build process of the NSC. Configuring the uIP stack is done via modifications of the `contiki-conf.h` file (refer to listing 3) before compilation time, as it is custom for Contiki. To test different uIP configuration options, it is necessary to recompile the shared library.

Listing 3. uIP configuration options - excerpt from `contiki-conf.h`
```
#define UIP_CONF_UDP                1
#define UIP_CONF_MAX_CONNECTIONS    40
#define UIP_CONF_MAX_LISTENPORTS    40
#define UIP_CONF_BYTE_ORDER         UIP_LITTLE_ENDIAN
#define UIP_CONF_TCP                1
#define UIP_CONF_TCP_SPLIT          0
#define UIP_CONF_LOGGING            0
#define UIP_CONF_UDP_CHECKSUMS      1
#define UIP_CONF_BUFFER_SIZE        400
#define PACKETBUF_CONF_SIZE         400
...
// definitions of Contiki netstack components
#define NETSTACK_CONF_MAC           nsc_mac_driver
#define NETSTACK_CONF_RDC           nullrdc_driver
#define NETSTACK_CONF_RADIO         nullradio_driver
#define NETSTACK_CONF_FRAMER        framer_nullmac
#define NETSTACK_CONF_NETWORK       uip_driver
```

The customized network stack (netstack) components are a peculiarity of Contiki. Contiki's network stack allows to exchange modules and drivers like the MC protocol, the framer or the radio driver. Customized drivers were introduced to interconnect the uIP stack and dependent modules with the simulation support code and the NSC.

Future releases of Contiki containing new and changed versions of uIP should work without modifications as long the basic uIP API is not changed and configuration options are reviewed for possible changes. We switched during our integration between the Contiki releases 2.6, 2.7, and the Github development version. The latter one required changes of the configuration options in the `contiki-conf.h` file and adaptations of the support code for the communication between the shared uIP library and OMNeT++ through the NSC. These changes were necessary because uIP was extended with a socket API to ease maintenance and code development inside Contiki. The support code was changed to use the new socket API of uIP. As soon as the new Contiki version 3.0 will be released, we plan to test the integration of this latest release of uIP into the NSC to determine if our integration process is valid for newer releases too.

The validation process of new stacks is suggested to consist of two steps [19]. In the initial testing, the stack should be able to perform basic operations and then be expanded to communication tests with multiple instances. The goal is to ensure that code extraction and application of the *globaliser* produced a functioning simulation model, which still needs to be fully validated [19]. The already supported stacks can be utilized for additional functionality and interoperability tests.

For the validation of a new network stack, a simulation and testbed comparison is suggested [19]. The NSC interface is able to trace and save packets in the `libpcap` file format. These packet traces can be directly compared to `tcpdump` traces from an equivalent real-world test setup. The goal is to achieve realistic behavior in the simulation compared to the real world in different use case scenarios.

### IV. APPLICATIONS AND USE CASES

The NSC has been proven to be useful for performance testing and has shown a wide variation between TCP/IP implementations and significant differences from simplified models [11]. The ability to simulate with real implementations can be used both as a validation and an analyzing tool. We will further discuss three use cases for both fields of applications.

As we are interested in simulations of complex Internet of Things scenarios, where Internet-based systems work together with sensor networks running with uIP, we first want to use OMNeT++ with the NSC to conduct interoperability tests between uIP and other real world TCP/IP stacks. As it was pointed out in section II, uIP has well-known weaknesses when interacting with full-scale TCP/IP stacks. Certain issues can be omitted or mitigated by selecting suitable configuration parameters for uIP. Examples here are the fragmentation limits, buffer sizes, and timing options for TCP and IP itself. An example IoT scenario for our evaluation is depicted in figure 2.

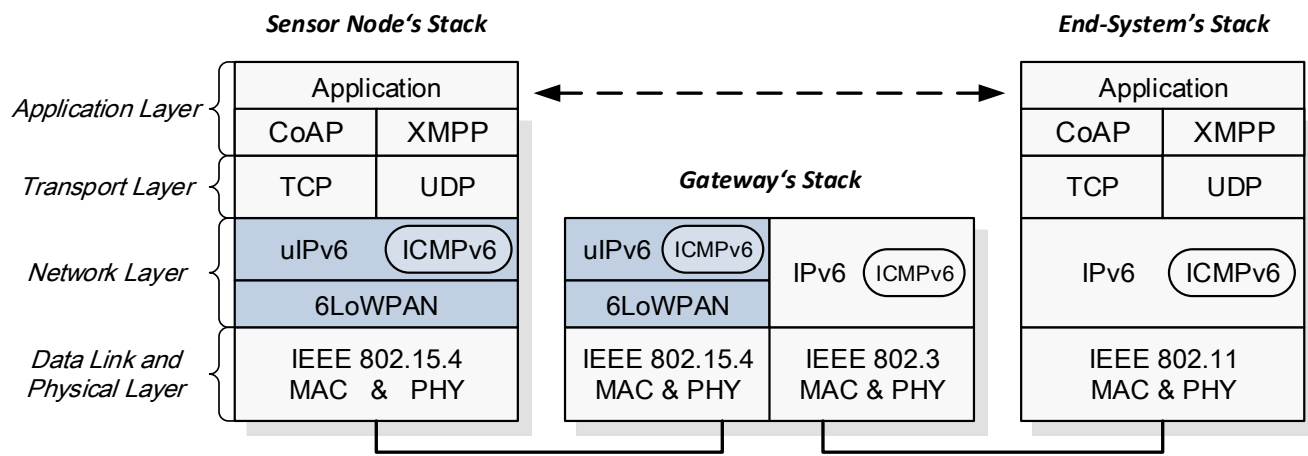

Fig. 2. Internet of Things Evaluation Scenario

In combination with our previously introduced 6LoWPAN simulation model for OMNeT++ [20], we want to evaluate the performance of both uIP and 6LoWPAN fragmentation when resource-constrained sensor nodes communicate with gateways and Internet-based systems (cp. figure 2). Fragmentation on the lower layers of the uIP stack is an interesting research field, as the exact fragmentation border depends on the actually used sensor node and its transceiver [21]. Due to different fragmentation limits, the supported IP payload sizes for uIPv4 / uIPv6, and in general for Contiki-based systems, vary for different sensor nodes. We want to review what influence fragmentation in combination with problems like the "one-packet-at-a-time" principle has on actual IoT application protocols. This goes beyond a simple performance analysis, as packet fragmentation can also affect protocol operations and timer behavior. The ability to trace the protocol operations with `libpcap` traces directly captured from the NSC will hopefully help us to identify issues here.

A third use case is the extensive testing of interoperability issues. The scenario from figure 2 does depict a case where a resource-constrained device communicates with a non-constrained device (i.e., a gateway in this case). Interoperability testing also needs to consider scenarios with heterogeneous sensor nodes. A study [8] of interoperability between Contiki's uIP and tinyOS' embedded IP stack *blip* [22] showed the need for performance tests and additional interoperability and compliance tests for all components of a TCP/IP stack. The authors of [8] discovered that the packet reception ratio decreased for as much as 10% when switching from homogeneous to heterogeneous deployments of sensor nodes with Contiki and tinyOS combined. We hope that future simulations of such heterogeneous scenarios will enable further insights on the actual performance of uIP when used in combination with other stacks.

To summarize, simulating scenarios with the NSC will generally allow us to perform rapid, controllable, and repeatable experiments, which is why we strive to extend the NSC with uIP support to enhance OMNeT++-driven IoT simulations.

## V. ONGOING AND FUTURE WORK

We are currently in the testing phase (cp. section III) for the integration of uIPv4 into the NSC. A thorough testing of the code extraction and *globaliser* operations is in progress. Next steps will be a validation with the help of a real-world testbed based on several sensor nodes running Contiki applications. The integration of the IPv6 part of uIP is also currently in progress. Further testing of uIPv6 will follow once the uIPv4 integration is finalized and validated.

A fitting evaluation of our integration approach will be possible when the upcoming Contiki version 3 is released. Updates to the uIP internals are planed for this release. We hope that this future release can be used with the extended NSC without any code changes (with the exception of added and changed configuration parameters).

Future work could also include the integration of additional stacks from the Internet of Things field into the NSC and thus into OMNeT++. There are several other stacks available, from tinyOS's blip stack [22] to the stacks of RIOT [23] and other WSN operating systems, which could be reviewed for further integration and testing with the NSC.